\documentstyle[preprint,revtex]{aps}
\begin{document}
\draft
\preprint{IMSc/93-01}
\begin{title}
Classical Dynamics of Anyons and the Quantum Spectrum
\end{title}

\author{G.Date and M. V. N. Murthy}

\begin{instit}
Institute of Mathematical Sciences,\\
Madras 600 113, India.\\

\end{instit}

\begin{abstract}
In this paper we show that (a) all the known exact solutions of the
problem of N-anyons in oscillator potential precisely arise from the
collective degrees of freedom, (b) the system is pseudo-integrable ala
Richens and Berry.  We conclude that the exact solutions are trivial
thermodynamically as well as dynamically.

\end{abstract}

\pacs{PACS numbers~:~~03.20.+i, ~46.10.+z  }

\narrowtext
\section{Introduction}

Non-relativistic quantum mechanics in two space dimensions admits
the possibility of fractional statistics and particles obeying
fractional statistics are known as ``anyons''.\cite{lm,wil}
Anyons are objects
defined in the quantum framework described by a Lagrangian of the
form,
\begin{equation}
L=\frac{1}{2}\sum_{i=1}^{N}\dot{\vec r}_i^2 + \alpha \sum_{i<j}^{N}
\dot{\theta}_{ij} -V(\vec r_i);\hspace{1cm}
\theta_{ij}=tan^{-1}\frac{y_i-y_j}
{x_i-x_j},
\end{equation}
where $\vec r_i$ denote particle coordinates and $V(\vec r_i)$ is some
confining potential which we choose to be the harmonic oscillator
potential $V(\vec r_i)= \sum_{i=1}^N \vec r_i^2$.  The $\alpha$
dependent term is the statistical interaction.   The harmonic
oscillator
potential is convenient since the dynamics of the system is well
understood in the limit $\alpha = 0$ .  We may choose to describe the
system of anyons either through multivalued wavefunctions which occur
naturally in the quantum mechanics on multiply connected spaces or
equivalently interms of single valued wave functions in the presence
of statistical interaction.  Quantum mechanically a system of N-anyons
confined in a harmonic oscillator potential is an exceptional system
since many exact solutions to the energy eigenvalue problem are known
even when the many body Hamiltonian is
nonseparable.\cite{wu,chou,app,dunne,basu}

The known results about the spectrum of N-anyons confined in oscillator
potential fall into two categories: (a) Exact eigenvalues
which are linear functions of the statistical parameter $\alpha$,
$E_{jm} =2m+|j-\alpha \frac{N(N-1)}{2}|+N$, where j and m denote the
angular and radial quantum numbers, and (b) eigenvalues that are
nonlinear functions of $\alpha$   as evidenced by numerical
calculations for $N=3,4$ anyons\cite{sporre1,murthy1,sporre2}
and also meanfield calculations for
large $N$\cite{li}.  The special case in this context is $N=2$ where the
spectrum is completely subsumed by the category (a).
The nonlinear spectrum displays many
level crossings, some of which are of the Landau-Zener type(avoided
crossings).\cite{sporre1,murthy1,sporre2}
This has lead to the conjecture that the anyon system may
be nonintegrable\cite{sporre2} ( or even chaotic) for $N>3$.

Two comments are in order here: Firstly, what is the reason for the
existence of exact solutions of the type (a) in this nonseparable many
body system?  Infact the approach of Basu etal\cite{basu}
shows that j=0 exact
solutions form a subspace of the full Hilbert space which block
diagonalises the Hamiltonian.  This indicates that the system may be
atleast
partially separable. The origin of this has not been elaborated
yet to the best of our knowledge.  The second comment pertains to the
``integrability'' of the system.  If one considers the classical
Lagrangian, the statistical interaction is a total derivative.  The
Euler-Lagrange equations of motion are therefore the same as that of
an $2N$-dimensional oscillator and this ofcourse is an integrable
system.  Why do then the numerical results suggest nonintegrability?

In this paper we elaborate on and answer the above questions.  In
section 2, we briefly mention the known exact solutions to the quantum
many body problem.  In section 3 we outline the classical Lagrangian
formulation of N-anyons and prove the system is ``integrable'' in the
Liouville sense, ie., it admits $2N$
constants of motion in involution.  In section 4 we exhibit a choice
of coordinates at classical level which gives partial separability.
{}From this we can isolate a set of collective coordinates whose
semiclassical quantisation leads to the known exact solutions.  These
solutions correspond to the periodic orbits in the phase space of the
oscillator which continue to remain periodic even in the presence of
anyonic interaction. In section 5 we consider the integrability aspect
of the system.  We argue that even though we have $2N$ constants of
motion in involution there exist invariant surfaces which do not have
the topology of $2N$ dimensional torus.  Following
Berry-Richens\cite{richens} we conclude that the many anyon system
confined in oscillator potential is pseudointegrable.

\newpage

\section{Exact Solutions}
The quantum mechanical Hamiltonian corresponding to the Lagrangian in
Eq.(1) is given by,
\begin{equation}
H =[\frac {1}{2}\sum_{i=1}^{N}{p}_{i}^{2} +
\frac {1}{2}\sum_{i=1}^{N} r_{i}^{2}
-\alpha\sum_{j>i=1}^{N}\frac{\ell_{ij}}{r_{ij}^2}
+\frac{\alpha^{2}}{2}\sum_{i\neq j,k}^N\frac{\vec{r}_{ij}.\vec{r}_{ik}}
{r_{ij}^{2}  r_{ik}^{2}}],
\end{equation}
where
\[\ell_{ij}= (\vec{r}_{i} - \vec{r}_{j})\times(\vec{p}_{i}-\vec{p}_{j}).\]
and all distances have been expressed in units of $1/\sqrt {m\omega}$,
where $m$ is the mass of the particle and $\omega$ is the oscillator
frequency.
Notice that the statistical interaction is independent of the centre
of mass. We now briefly discuss the class of exact solutions already
known.\cite{wu,chou,app,dunne,basu}
For discussing the known class of exact solutions it is convenient
to use the complex coordinates $z_i, \bar z_i$ in terms of which
the Hamiltonian takes the form,
\begin{equation}
H = -2\sum_i \partial_i \bar \partial_i + \frac{1}{2} \sum_i
z_i \bar z_i
- \alpha \sum_{i<j}(\frac {\partial_{ij}}{\bar z_{ij}} -
\frac {\bar \partial_{ij}}{ z_{ij}}) +\frac {\alpha^2}{2}
\sum_{i \ne j,k} \frac {1}{\bar z_{ij}z_{ik}},
\end{equation}
where $\partial_i = \partial /\partial z_i; \partial_{ij}=\partial_i
-\partial_j$, etc.,
and the eigenvalue equation is,
\begin{equation}
H\psi(z_i,\bar z_i) = E\psi(z_i,\bar z_i).
\end{equation}
The conserved angular momentum $J$ with eigenvalues denoted by $j$,
is given by,
\begin{equation}
J=z_i\partial_i -\bar z_i \bar \partial_i.
\end{equation}
Defining variables,
\begin{equation}
X = \prod_{i<j}^N z_{ij} ; \hspace{1cm} t=\sum_{i=1}^N z_i \bar z_i,
\end{equation}
we may classify the known exact solutions as follows:

(a)$j < 0 $ solutions-
\begin{equation}
\psi_{j} = |X|^{\alpha}\phi_{j}(\bar z_i)e^{-t/2} ;\hspace{0.5cm} j < 0
\end{equation}
with the energy eigenvalues given by,
\begin{equation}
E = N - j + \alpha \frac {N(N-1)}{2}.
\end{equation}
Here and in what follows $\phi_j$ generically denotes an eigenfunction
of the angular momentum operator $J$ with eigenvalue $j$. For
elucidating the solutions the specific form of $\phi$ is irrelevent.

(b) $j = 0$ solutions-
\begin{equation}
\psi_0 = |X|^{\alpha}\phi_{0}(t)e^{-t/2} ;\hspace{0.5cm} j = 0.
\end{equation}
with $\phi_0(t) (=\sum_{k=1}^m C_k t^k)$ a polynomial of degree m in t.
The corresponding energy eigenvalues are given by,
\begin{equation}
E = N + 2m + \alpha \frac {N(N-1)}{2},
\end{equation}
The second solution is necessarily bosonic since $t$ is symmetric,
where as the first solution
needs explicit symmetrization and antisymmetrization of the wavefunction
in terms of $\bar z_i$ to obtain the bosonic and fermionic wavefunctions.
Since this is always possible, the degeneracy of the first type solution
is exactly the same for both bosonic and fermionic type solutions for any
given angular momentum $j$ ($ < 0 $). We may also take a combination
(product) of the solutions the types
discussed above, to get further $j<0$ solutions. This then generates a
infinite tower radial excitations for each value of j.

(c) $j > 0$ solutions-
\begin{equation}
\psi_{j} = |X|^{-\alpha}\phi_{j}(z_i)e^{-t/2} ;\hspace{0.5cm} j > 0
\end{equation}
with the energy eigenvalues given by,
\begin{equation}
E = N + j - \alpha \frac {N(N-1)}{2}.
\end{equation}
Caution must be excercised in choosing the value of $j$  for these
solutions since the wave function is not square integrable for all
values of $j$.  Infact the lower bound is obtained requiring that the
wave function be square integrable over the whole domain of $\alpha
(0\le \alpha \le 1)$. This means that $j > (N-1)(N-2)/2$.
If however this condition is not satisfied then the wave function
remains regular only for some values of $\alpha$
($0\le \alpha \le 2j/N(N-1)$)
but not for all $0 \le \alpha \le 1$ which gives rise to the so called
noninterpolating solutions which have also been discussed in the
literature.\cite{murthy2}

All these solutions for the energy eigenvalues have a linear
dependence on $\alpha$ with a coefficient $\pm N(N-1)/2$ while
the corresponding eigenfunctions
are finite order polynomials apart from the overall $\mid X \mid^{\pm
 \alpha}$
and the Gaussian factors. These solutions (a)-(c) cover all the known
exact solutions.  However it is by now known that these exact
solutions form only a subset of the full Hilbert space and existence
of nonlinear solutions has been shown numerically as well as through
meanfield calculations.  It is our aim here to understand the reason
for the existence of this dichotomy.

\section{Constants of Motion}
We begin with the analysis of the classical Lagrangian given by
Eq.(1). It is convenient to write the Lagrangian in the form,
\begin{equation}
L=\frac{1}{2}\sum_{i=1}^{N}[\dot{\vec r}_i^2-\vec r_i^2]
+ \alpha \sum_{i<j}^{N} \frac {\vec r_{ij} \times \dot {\vec r_{ij}}}
{\vec r_{ij}.\vec r_{ij}},
\end{equation}
where the dots indicate the time derivatives.
The first step is to introduce relative coordinates and
separate the trivial centre of mass degree of freedom. To this end we
write,
\begin{equation}
\vec{\rho}_a=[\frac{1}{\sqrt{a(a+1)}}\sum_{k=1}^{a}
\vec{r}_k-\sqrt{\frac{a}{a+1}}\vec{r}_{a+1}];~~~a=1,...,N-1
\end{equation}
with inverse relation given by,
\begin{equation}
\vec{r}_i=[-\sqrt{\frac{i-1}{i}}\vec{\rho}_{i-1}
+\sum_{k=i}^{N-1}\frac{\vec{\rho}_k}{\sqrt{k(k+1)}}] +\vec{R}_{cm} \equiv
A_{i}^{a}\vec {\rho}_{a}+\vec{R}_{cm},
\end{equation}
where $\vec{\rho}_{a}, a=1,...,N-1$ are dimensionless relative coordinates and
$\vec{R}_{cm}$ is the centre of mass coordinate.
It follows that,
$$ \sum_i A_i^a =0; \hspace{1cm} \sum_i A_i^a A_i^b = \delta^{ab}.$$
It is straight forward to see that,
$$ L=L_{CM}+L_{rel},$$
where
$$ L_{CM}=\frac{1}{2}[\dot {\vec R}_{CM}^2 -\vec R_{CM}^2],$$
$$ L_{rel}=\frac{1}{2}[\dot {\vec \rho_a^2}-\vec \rho_a^2] +
\alpha \sum_{i<j} \frac{A_{ij}^a A_{ij}^b \vec \rho_a \times \dot{ \vec
\rho}_b}{A_{ij}^c A_{ij}^d \vec \rho_c . \vec \rho_d},$$
where $A_{ij}^a = A_i^a -A_j^b$.  From now on we concentrate only on
the $L_{rel}$ and drop the subscript.  It is easy to see that the
Euler-Lagrange equations of motion are,
\begin{equation}
\ddot{\vec \rho}_a = -\vec \rho_a.
\end{equation}
There is no $\alpha$ dependence in these equations since the
corresponding term is a total derivative.  From these equations it
follows that
\begin{equation}
{\cal E}_a \equiv \frac{1}{2} [\dot{\vec{\rho}}_a^2+\vec{\rho} _a^2];
\hspace{0.5cm} l_a=\vec{\rho} _a \times \dot{\vec{\rho}}_a; \hspace{0.5cm}
a=1,...,N-1
\end{equation}
are constants of motion.  In the Hamiltonian formulation the $\dot
{\vec{\rho}}_a$ are expressed in terms of conjugate momenta and coordinates
which do contain the $\alpha$ dependence,
\begin{eqnarray}
P_{ax}=\frac{\partial L}{\partial \dot{\rho}_{ax}}=\dot{\rho}_{ax}
-\alpha {\cal A}_{ax}; \hspace{0.5cm}
{\cal A}_{ax}=\sum_{i<j} \frac{A_{ij}^a A_{ij}^b \rho_{by}}
{A_{ij}^c A_{ij}^d \vec \rho_c . \vec \rho_d},\\
P_{ay}=\frac{\partial L}{\partial \dot{\rho}_{ay}}=\dot{\rho}_{ay}
+\alpha {\cal A}_{ay}; \hspace{0.5cm}
{\cal A}_{ay}= \sum_{i<j} \frac{A_{ij}^a A_{ij}^b \rho_{bx}}
{A_{ij}^c A_{ij}^d \vec \rho_c . \vec \rho_d},
\end{eqnarray}
and the constants of motion in relative coordinates are,
\begin{eqnarray}
{\cal E}_a = \frac{1}{2}[(P_{ax}+\alpha {\cal A}_{ax})^2
+(P_{ay}-\alpha {\cal A}_{ay})^2 +\vec {\rho}_a^2]\\
l_a = \vec{\rho}_a\times \vec{P}_a -\alpha (\rho_{ax}{\cal A}_{ay}
+\rho_{ay}{\cal A}_{ax}).
\end{eqnarray}
Clearly the Hamiltonian is given by
$$ H = \sum_{a=1}^{N-1} {\cal E}_a.$$
Note that the $\alpha$-dependent term in $L$ is well defined only if
$\vec{r}_{ij}\ne 0$ for all i,j.  That is only if
$$A_{ij}^a A_{ij}^b\vec{\rho}_a.\vec{\rho}_b ~\ne~ 0 ~~\forall~~ i,j$$
Consequently the expressions
for ${\cal E}, l_a$ and $H$ are also valid only if $\vec{r}_{ij}$ is
not zero.  In effect the classical configuration space on which $L$ is
well defined is the space
\begin{equation}
Q^{N-1} = R_2^{N-1}-\Delta; \hspace{0.5cm} \Delta = \{~\vec{\rho}_a~/
{}~A_{ij}^c A_{ij}^d~ \vec{\rho}_c.\vec{\rho}_d~ =~0 ~~
for~ some~ pair(s)~~i,j\}.
\end{equation}
The space $Q^{N-1}$ is not simply connected.  Its fundamental
group is the same as the fundamental group of $\{R_2 - (N-1)~~
points \}$ which is known to be nontrivial.  For $N=2$, $\pi _1 =
Z$ while for $N \ge 3,~~ \pi_1$ is nonabelian.

The corresponding phase space is the cotangent bundle of Q on which
${\cal E}_a, l_a$ and $H$ are well defined.  This phase space is
topologically nontrivial and for our purposes we do not need the full
machinery for handling this topologically nontrivial phase space. It
is sufficient to pretend that the full space is still $R^{N-1}_2$ but
simply avoid coincident points.

It is straight forward then to prove the following Poisson bracket
relations,
\begin{eqnarray}
\{{\cal E}_a,{\cal E}_b\} & = &\{l_a,l_b\} = \{{\cal E}_a,l_a\}=0
\hspace{0.3cm} \forall \hspace{0.3cm} a,b=1,...,N-1, \\
\{H,{\cal E}_a\}&  = & \{H,l_a\} =  0.
\end{eqnarray}
Thus for $4(N-1)$ degrees of freedom $P_a, \rho_a$ we have
$2(N-1)$ constants of motion in involution.  So a necessary condition
for a system to be integrable is satisfied.  We will tentatively refer
to this system as being integrable (or potentially integrable) in the
``Liouville sense''.

\section{Collective Modes}

The anyonic interaction term in $L$ is invariant under two sets of
time independent transformations:
\begin{eqnarray}
\vec{\rho}_a &\rightarrow & \vec{\eta}_a = R(\theta)\vec{\rho}_a
\hspace{0.3cm} \forall \hspace{0.3cm} a,\\
\vec{\rho}_a &\rightarrow & \vec{\eta}_a = \lambda\vec{\rho}_a
\hspace{0.3cm} \forall \hspace{0.3cm} a,
\end{eqnarray}
where $R(\theta)$ denotes a rotation of the vector by an angle
$\theta$ in two dimensions.
Therefore for any given configuration $\{\vec{\rho}_a\}$ at any given
time t we can always rotate the axes so that  $\rho_{1y}=0$ (say).  Thus
by making a time dependent rotation we can ensure that $\eta_{1y} =0$
for all t.  Therefore define,
\begin{equation}
\vec{\rho}_a = \pmatrix { cos\theta & -sin\theta \cr sin\theta &
cos\theta } \vec{\eta}_a,
\end{equation}
The following identities are easy to prove:
\begin{eqnarray}
\vec {\rho}_a.\vec {\rho}_b & = & \vec {\eta}_a.\vec {\eta}_b ,\\
\vec {\rho}_a \times \vec {\rho}_b & = &
\vec {\eta}_a \times \vec {\eta}_b ,\\
\dot{\vec {\rho}}_a^2 & =& \dot{\vec {\eta}}_a^2+2\dot\theta \vec {\eta}_a
\times \dot{\vec {\eta}}_a+\dot{\theta}^2\vec {\eta}_a^2,\\
\vec {\rho}_a \times \dot{\vec {\rho}}_b & = &
\dot{\theta}\vec {\eta}_a.\vec {\eta}_b + \vec {\eta}_a \times
\dot{\vec {\eta}}_b.
\end{eqnarray}
Now the set $\{\eta_a ; a=1,...,N-1\}$ is effectively a $2(N-1)-1$
dimensional vector since $\eta_{1,y}=0$ for all time t.  We can
therefore introduce the standard spherical coordinates by the usual
procedure,
\begin{equation}
\vec{\eta}_a \equiv R \vec{\xi}_a; \hspace{0.3cm}\xi_{1y}=0;
\hspace{0.3cm} \sum_a \xi_a^2 = 1,
\end{equation}
where
\begin{eqnarray}
\xi_{N-a,x} & = &s_1 s_2 ...s_{2a-2}c_{2a-1},\\
\xi_{N-a,y} & = &s_1 s_2 ...s_{2a-1}c_{2a},\\
\xi_{1,x} & = &s_1 s_2 ...s_{2N-4},\\
\xi_{1,y} & = & 0,
\end{eqnarray}
where $a=2,...,N-1$ and $s_{\mu}=\sin \theta_{\mu},
c_{\mu}=\cos \theta_{\mu}$.
Interms of these variables the Lagrangian may be
rewritten as
\begin{equation}
 L=\frac{1}{2}[\dot {R}-R^2+R^2\dot{\theta}^2+2R^2
\dot{\theta}\sum_a\vec{\xi}_a \times \dot{\vec {\xi}}_a+R^2 \sum_a
\dot {\vec{\xi}}_a^2 ] + \alpha \frac{N(N-1)}{2}\dot {\theta}+
\alpha \sum_{i<j} \frac{A_{ij}^a A_{ij}^b \vec \xi_a \times \dot{\vec
\xi}_b}{A_{ij}^c A_{ij}^d \vec \xi_c . \vec \xi_d},
\end{equation}

Since $\vec\eta_a$ are obtained from $\vec\rho_a$ by the same rotation
matrix for all a, the angle $\theta(t)$ clearly describes a collective
rotation of all the N-anyons about the centre of mass(say).  The anyonic
interaction term ( a total time derivative) is manifestly independent
of $R(t)$. This may also be regarded as a collective mode as discussed
below.  It is the semiclassical quantisation of these two modes that
yields the exactly known energy eigenvalues.   To elaborate these
points let us consider the Euler-Lagrange equations of motion.  Since
the $\alpha$ dependent part of the Lagrangian is a total time
derivative we can ignore it for analysing the classical equations of
motion.  Clearly these equations are identical to that of the
oscillator equations of motions as given in Eq.(16).  Translating this
into the spherical coordinates we obtain,
\begin{equation}
[\ddot{R}+R-R\dot{\theta}^2]\vec{\xi}_a-[R\ddot{\theta}+2\dot{R}
\dot{\theta}]V\vec{\xi}_a+2[\dot{R}-R\dot{\theta}V]\dot{\vec{\xi}}_a
+R\ddot{\vec{\xi}}_a =0,
\end{equation}
where
$$V=\pmatrix{0 & 1 \cr -1 & 0}$$
and in the matrix equation above the $\vec{\xi}$ is a column vector
with two elements.  The matrix V between two vectors essentially
generates their cross product.  Taking the dot product with
$\vec\xi_a$ and summing over $a$ we find,
\begin{equation}
[\ddot{R}+R-R\dot{\theta}^2]=2R\dot{\theta}\sum_a\vec{\xi}_a \times
\dot{\vec{\xi}}_a +R\sum_a\dot{\vec{\xi}}_a^2 ,
\end{equation}
where we have made use of the identities,
$$ \sum_a \vec{\xi}_a.\dot{\vec{\xi}}_a=0, \hspace{0.3cm} \sum_a
\vec{\xi}_a.\ddot{\vec{\xi}}_a=-\sum_a \dot{\vec{\xi}}_a^2$$.
Taking the cross product with
$\vec\xi_a$ and summing over $a$ we find,
\begin{equation}
[R\ddot{\theta}+2\dot{R}\dot{\theta}]=-2\dot{R}\sum_a
\vec{\xi}_a \times
\dot{\vec{\xi}}_a -R\sum_a\vec{\xi}_a \times
\ddot{\vec{\xi}}_a.
\end{equation}
Using the above two relations Eq.[38] can be rewritten as,
\begin{equation}
R\ddot{\vec{\xi}}_a
+2[\dot{R}-R\dot{\theta}V]\dot{\vec{\xi}}_a
+[2R\dot{\theta}\vec{\xi}_b \times
\dot{\vec{\xi}}_b +R\dot{\vec{\xi}}_b^2]\vec{\xi}_a+
[2\dot{R}\vec{\xi}_b \times
\dot{\vec{\xi}}_b +R\vec{\xi}_b \times
\ddot{\vec{\xi}}_b]V\vec{\xi}_a
=0,
\end{equation}
where sum over $b$ is assumed.
Here the equations motion for $R$ and $\theta$ are still coupled to
all the other internal coordinates and as yet they are not collective
coordinates.  We therefore  need to impose a set of initial conditions
which
may lead to the separation of these two modes as collective modes from
the internal variables $\theta_{\mu}$.  To realise this let at time
$t=0$, all velocities $\dot{\vec{\xi}}_a(t=0) =0$ for all $a$.  Then the
above equations reduce to,
\begin{eqnarray}
R[\ddot{\vec{\xi}}_a +(\sum_b \vec{\xi}_b \times
\ddot{\vec{\xi}}_b) V\vec{\xi}_a]=0, \\
\ddot{R}+R-R\dot{\theta}^2=0, \\
R\ddot{\theta}+2\dot{R}\dot{\theta}=-R\sum_b \vec{\xi}_b \times
\ddot{\vec{\xi}}_b.
\end{eqnarray}

Now consider the first equation for $R\ne 0$ and $a=1$.  From the initial
conditions it is obvious that,
$$\ddot{\xi}_{1x}=0; \hspace{0.5cm}\xi_{1x}
\sum_b (\vec{\xi}_b\times\ddot{\vec{\xi}}_b) = 0$$
because $\xi_{1y}=0 ~~forall~~ t$.

Now {\it if} $\xi_{1x}$ is nonzero, then
$\sum_b\vec{\xi}_b \times\ddot{\vec{\xi}}_b=0,$ and hence
$$\ddot{\xi}_b=0~~\forall~~b.$$
Since the equations of motion are second order in t, we have proved that:
{\it if} $\dot{\vec{\xi}_a}(0) = 0 ~~\forall~~ a ,~~ and~~ R(0) ,~
\xi_{1x}(0) $ are
non zero {\it then} $ \forall~~ t $,
$$\vec{\xi_a}(t) = \vec{\xi_a}(0) , $$
$$\ddot{R} + R - R {\dot{\theta}^2} = 0, $$
$$R {\ddot{\theta}} + 2 {\dot{R}} {\dot{\theta}} = 0.$$
For oscillator ($\alpha = 0$) R(0) and $\xi_{1x}(0) $ being non zero is
a special class of initial condition.
Indeed in general for a second order equation system if both first and
second order derivatives vanish at any t then this can hold only for
some restricted class of coordinate values. For the equations of motions
considered in the configuration space $R^{N-1}_2$ the restricted class
is precisely characterised by $ R(0) \ne 0 $ and $ \xi_{1x}(0) \ne 0 $
However when $\alpha \ne 0$ the
configurations space is $Q^{N-1}$ which does not contain points which
have R = 0 and $\xi_{1x} = 0 $. Thus there are no restrictions on the
initial conditions in $Q^{N-1}$.

The above initial conditions amount to freezing the ``internal motion'' of
the anyons. The remaining motion is a collective motion described by
R(t) and $\theta(t)$ which is just a $2(N-1)$ dimensional oscillator. R(t)
being nonzero implies that the angular momentum must be nonzero. This
explains in what sense R(t) can be interpreted as a collective mode.
For describing the motion of collective mode the effective Lagrangian
is
\begin{equation}
L_{eff}= \frac{1}{2} [\dot{R}^2 + R^2 \dot{\theta}^2 -R^2] +\alpha
\frac{N(N-1)}{2} \dot{\theta}.
\end{equation}
This is identical to the Lagrangian in  the relative coordinate for a
two anyon system with $\alpha \rightarrow \alpha \frac{N(N-1)}{2}$
and in $2(N-1)$ dimensions.
Semiclassical quantisation will then reproduce all exactly known
energy eigenvalues summarised in Sect.2.
In this effective Lagrangian picture the only
known memory of N resides in the coefficient of $\alpha$.  This can be
understood by noting that when all the N-anyons are rotated by $2\pi$
about the centre of mass they also circle each other to pick up the
extra phase.

If on the otherhand  we consider
$\dot{R}= \dot{\theta}=0$ at $t=0$ class of initial
conditions then we see that $\ddot{R}, \ddot{\theta}$ are nonzero and
hence $R(t), \theta(t)$ do depend on $\theta_{\mu}$.  Thus the
collective motion is not fully decoupled from the ``internal motion''.
We therefore refer to this system as partially separable.

Incidentally the same conclusions can be drawn from the Hamiltonian
formulation. For completeness we give the relevent expressions and
arguements.  The conjugate momenta for the hyperspherical variables
are given by,
\begin{eqnarray}
P_R & = & \dot{R},\\
P_{\theta}& = & R^2 \dot{\theta} + R^2 \sum_{\mu}
\dot{\theta}_{\mu}F_{\mu} +\alpha
\frac{N(N-1)}{2}, \\
P_{\mu} & = & R^2 r_{\mu}^2 \dot{\theta}_{\mu} +
R^2 \dot{\theta}F_{\mu} +\alpha G_{\mu},
\end{eqnarray}
where,
$$ r_{\mu}=s_1 s_2...s_{\mu-1},~~\mu=1,...,2N-3,~~r_1=1$$
and $F_{\mu}$ and $G_{\mu}$ are defined through,
$$ \sum_{\mu =1}^{2N-4}\dot{\theta}_{\mu}F_{\mu} \equiv
\sum_{a=1}^{N-1}\vec{\xi}_a  \times \dot{\vec {\xi}}_a$$
$$ \sum_{\mu =1}^{2N-4}\dot{\theta}_{\mu}G_{\mu} \equiv
\sum_{i<j} \frac{A_{ij}^a A_{ij}^b \vec \xi_a \times \dot{\vec
\xi}_b}{A_{ij}^c A_{ij}^d \vec \xi_c . \vec \xi_d}.$$
The Hamiltonian is then given by
\begin{equation}
H=H_1+H_2,
\end{equation}
where
\begin{equation}
H_1 = \frac{1}{2} [P_R^2 + R^2] + \frac{(P_{\theta}-\alpha
\frac{N(N-1)}{2})^2}{2R^2},
\end{equation}
\begin{equation}
H_2 = \frac{1}{2R^2} [ \sum_{\mu} \frac{(P_{\mu}-\alpha
G_{\mu})^2}{r_{\mu}^2} +\frac{(P_{\theta}-\alpha
\frac{N(N-1)}{2}-\sum_{\mu}\frac{F_{\mu}(P_{\mu}-\alpha
G_{\mu})}{r^2_{\mu} })^2}{1-\sum_{\mu}\frac{F_{\mu}^2}{r_{\mu}^2}}
-(P_{\theta}-\alpha
\frac{N(N-1)}{2})^2].
\end{equation}
It is easy to see that for the special initial conditions
$\dot{\theta}_{\mu} = 0 $, $H_2 =0$.  Also the Poisson bracket
$\{H_1,H_2\} \propto H_2$
vanishes for these initial conditions.
The corresponding quantum statement would $[H_1,H_2] \propto H_2$,
and therefore
if we consider the
subspace $\{\psi\}$ on which $H_2\psi =0$
then $H_1$ will act invariantly on this subspace.
Further $H=H_1$ on this subspace and thus the
eigenstates of $H_1$ will be exact eigenstates of the full
system and conversly. But the
problem of solving $H_1$ is analogous to that of two anyon problem
with $\alpha \rightarrow \alpha \frac{N(N-1)}{2}$.
Therefore these solutions
are given by,
\begin{eqnarray}
E & = & 2m + |j - \alpha \frac{N(N-1)}{2}| + (N-1), \\
\psi & = & C R^{|j-\alpha|}\exp^{-R^2/2}f(R),\\
H_1 \psi & = & E \psi
\end{eqnarray}
where $f(R)$ is some polynomial of degree $2m$ in $R$.  In general the
normalisation constant C in
$\psi$ will have dependence on $\theta_{\mu}$'s, restricted by
$H_2\psi=0$.   These are
precisely all the known solutions and the above arguement shows that
there are no more solutions  satisfying $H_1 \psi = E \psi $ and $H_2
\psi = 0$.  Thus $H_2 \psi = 0 $ characterises the subspace spanned by
all the known exact solutions.  So the quantum counterpart of
classical partial separability implies the existence of these
solutions.

\section{Integrability}

We now come back to the integrability aspect. In Sect.3 we exhibited a
set of $2(N-1)$ constants of motion and regarded the system as
potentially integrable.
However for the system to be integrable via
action angle variables further conditions have to be
satisfied\cite{arnold}:  If
$M({\cal E }_a,l_a)$ denote the set of points in phase space at which
the constants of motion have values ${\cal E }_a,l_a$ then (a) $M$ is
a $2(N-1)$ dimensional submanifold iff ${\cal E }_a,l_a$ are all
independent, (b) if $M$ is a submanifold which is compact and
connected  then $M$ is
a $2(N-1)$ dimensional torus. If these conditions are satisfied then
one can introduce action-angle variables in the neighbourhood of $M$.
The orbits on $M$ will all be conditionally periodic in general and
for the periodic orbits  one may use Bohr-Sommerfeld
quantisation to get a subset of eigenvalues.

We also saw in the previous  section that there exist  $M({\cal E
}_a,l_a)$'s for which action-angle variables can be introduced.  These
correspond to the collective motion of N-anyons and their
semiclassical quantisation will reproduce the exactly known
eigenvalues described earlier. The Hamiltonian in Eq.(50)
being identical in structure to the oscillator Hamiltonian, the
orbits of this Hamiltonian will be in one-to-one correspondence with the
orbits of the oscillator and hence they will all be periodic.  However
if we go away from these initial conditions some of the orbits of
oscillator will cease to be periodic when $\alpha$ dependence is
introduced.  For instance the Euler-Lagrange equations
$$\ddot{\vec{r}}_i=-\vec{r}_i \Rightarrow
\ddot{\vec{r}}_{ij}=-\vec{r}_{ij} ~~(independent~ of~ \alpha).$$
For any pair $ij,~i\ne j$, $\vec{r}_{ij}$ in general describes an
ellipse in the configuration space.  However for orbital angular
momentum zero, the ellipse degenerates to a straight line, hence
$\vec{r}_{ij}=0$ is on such an orbit.
Since these points are not admissible when $\alpha
\ne 0$ these orbits cannot lead to periodic orbits in the phase space.
For all choices of ${\cal E}_a, l_a$ such that for some initial
conditions the phase space orbits will have $r_{ij}$ approaching
arbitrarily close to zero, $M({\cal E}_a,l_a)$ cannot be a torus. (M,
even if a manifold will fail to be compact and/or connected.)
Therefore for
$\alpha\ne 0$ although we have a potentially integrable system it is
not generically integrable via action-angle variables.
The existence of such an
$M$ indicates the possibility of  ``extreme sensitivity'' to the initial
conditions.  This is a possible reason for the ``level repulsions'' seen
numerically. This bears a resemblence to the billiard system
considered by Richens and Berry \cite{richens} though the reasons
for the failure of integrability via
action-angle coordinates appears to be different.
Following Richens and Berry we conclude that our system
is Pseudointegrable\cite{richens}.

\section{Summary and Conclusions}

In this paper we have exhibited two properties of the many anyon
system: (1) The partial separability of collective motion and (2)
pseudointegrability. We have shown that property (1) explains the
existence of the exactly known eigenvalues which is some what uncommon
for a generic many body problem.  {\it This also shows that the exactly
known spectrum incorporates only a some what trivial aspect of anyon
dynamics}. The nontrivial aspects although partially uncovered by
numerical results are still elusive.  Property (2) enables one to
understand  qualitatively the origin of the ``level repulsion'' seen in
some
numerical results.  Further study of pseudointegrability , for example
the topology of level sets M, may enable one to adapt different
techniques to get a handle on the nontrivial aspect of the spectrum.
We have demonstrated that the N-anyon system is  a nontrivial example
of a many body  pseudointegrable system
apart from the known problem of a billiard in polygonal enclosures with
a regular obstacle inside. In the light of the pseudointegrability, it
seems that the characterisation of quantum integrability must be
understood with further conditions , for example self-adjointness of
the constants of motion in involutions and the existence of common
dense domains.

We thank R.Ramaswamy for drawing our attention to the notion of
pseudointegrable systems and Radha Balakrishnan for discussions.

\end{document}